\title{ {\em Crypto-Battleships} or How to play Battleships game over the Blockchain?}
\author{
Guy Barshap \footnote{ E-mail: Barshag@post.bgu.ac.il } - BGU university of Israel. } 
\date{}
\documentclass[12pt]{article}
\usepackage{hyperref}

\usepackage{tikz}
\usepackage{amsmath}

\usetikzlibrary{shapes.geometric, arrows}

\tikzset{
    green/.style  = {draw, rectangle, 
      minimum width=1.6cm, minimum height=1cm, 
      outer sep=1mm,
      text centered, text width=1.2cm, font=\footnotesize,
      draw=black, fill=green!30}, 
    blue/.style   = {draw, rectangle,
      minimum width=8cm, minimum height=1cm,
      text centered, text width=5.0cm, font=\footnotesize,
      draw=black, fill=blue!30}, 
    blueM/.style   = {draw, rectangle, 
      minimum width=1.6cm, minimum height=1cm, 
      outer sep=1mm,
      text centered, text width=2.8cm, font=\footnotesize,
      draw=black, fill=white!30},       
    yellow/.style = {draw, rectangle, 
      minimum width=8cm, minimum height=1cm, 
      text centered, text width=5.0cm, font=\footnotesize,
      draw=black, fill=yellow!30}, 
}
\begin{document}
\maketitle

\begin{abstract}

Battleships is a well known traditional board game for two players which dates from World War I.
Though, the game has several digital version implementations, they are affected by similar major drawbacks such as fairness and a trust model that relies on third party.
In this paper, we demonstrate how to implement a fair, resistant to denial-of-service, where the honest winner earns the deposit money {\em immediately}. The game is built on a permissionless Blockchain that supports Turing complete smart-contract computation. Furthermore, we provide a full working game implementation\footnote{ The front-end game will be released  on the website: \url{http://www.playonchains.com}. 
 } of this proposition over the Ethereum Blockchain.

\end{abstract}

\section{Introduction - Basic Battleships rules}\label{Intro}
This section describes the basic rules of the classic battleships game, which establishes the basis for our novel  contribution. Battleships is a popular traditional board game\cite{wiki:battleships} for two players, where each player is required to hold a board game of $10 \times 10$ cell  size.
In each board, the players need to place down a fleet of 5 battleships, where each battleship has different sizes and occupies consecutive cells in the board. Each cell can either be a battleship's part or an empty cell. This classic game has the following stages:
\begin{enumerate}

\item {\bf Placing the Battleships.} 
To commence the game, each player needs to {\em secretly} arrange their fleets. The battleships can be placed in horizontal or vertical arrangement with each player possessing a predefined equal amount and types of battleships.

In this phase, each player also creates another board with the same size in order to record the torpedoes shot into the opponent's fleet, as well as their status (hit or miss).

\item {\bf Launching Torpedoes in turns.} This phase operates in rounds, where the players switch their roles in each round. In a single round, there is one player shooting a torpedo into one of the cells of the opponent's board and on announcing the exact cell that is being targeted, both players have to record the shot. Then, the opponent announces whether this cell contains a part of his own battleships (i.e the opponent announces whether it is a {\em hit} or {\em miss} shot).
In a situation where all parts of the ship have been affected, the owner of that ship must announce "This ship was sunk".

\item {\bf Termination of the game.}
In the eventual outcome that a fleet of one of the players is sunk (i.e all the battleships are sunk), the game ends and the opponent will be announced as the winner. Pedantic players at this stage will perform comparison of their own records against the private opponent's board arrangement to obtain some \footnote{This will not give them a full guarantee, since a malicious opponent could perform  {\em Dynamically changing battleship's location} attack. We describe that attack in Section \ref{attacks}} guarantee that they have not been cheated.

\end{enumerate}
\smallskip

  \begin{table}
  
  \begin{center}
  
\begin{tabular}{ |c||c|c|c|c|c|c|c|c|c|c| } 
 \hline
   & 1 & 2 & 3 & 4 & 5 & 6 &7 &8 & 9 & 10\\  \hline \hline
 A &  X & $\cdot$  & $\cdot$& $\cdot$ & $\cdot$& $\cdot$&$\cdot$ & $\cdot$& $\cdot$& $\cdot$\\ \hline
 B & X & $\cdot$  & $\cdot$& $\cdot$& $\cdot$& $\cdot$&$\cdot$ & $\cdot$& $\cdot$& $\cdot$\\  \hline

 C & X & $\cdot$  & $\cdot$&$\cdot$ &$\cdot$ &$\cdot$ &$\cdot$ & X &$\cdot$ &$\cdot$\\  \hline
 
 D &  $\cdot$ & $\cdot$  &X &X &X & X& $\cdot$&X  & $\cdot$& $\cdot$\\  \hline

 E & $\cdot$  &  $\cdot$ & $\cdot$& $\cdot$&$\cdot$ & $\cdot$& $\cdot$&  X & $\cdot$& $\cdot$\\ \hline

 F &  $\cdot$ & X & X &$\cdot$ & $\cdot$&$\cdot$ & $\cdot$& X & $\cdot$&$\cdot$\\  \hline

 G & $\cdot$  & $\cdot$  & $\cdot$& & X & $\cdot$ &$\cdot$ &  X & $\cdot$& $\cdot$\\  \hline
 
 H & $\cdot$  &  $\cdot$ & $\cdot$&$\cdot$ &$\cdot$ & $\cdot$& $\cdot$& $\cdot$& $\cdot$&$\cdot$ \\  \hline

 I &  $\cdot$ &  $\cdot$ & $\cdot$&$\cdot$ &$\cdot$ &$\cdot$ &$\cdot$&$\cdot$ &$\cdot$ & $\cdot$\\  \hline
  
 J & $\cdot$  & $\cdot$  &$\cdot$ &$\cdot$ & $\cdot$& $\cdot$&$\cdot$ &$\cdot$ & $\cdot$& $\cdot$\\  \hline

\end{tabular}
 \caption{A typical Battleships board game of $10 \times 10$ cell size and 5 battleships of sizes: $1\times 1$, $1\times 2$, \ldots , $5 \times 5$.}

  \end{center}

  \end{table}

\section{Why play battleships over the Blockchain?}

\subsection{Limitation of the Battleships' centrelized variant}\label{Centrelized}

A typical battleships game is normally hosted on a third party centralized server, however, this approach suffers from the following limitations.

{\bf Trusting the server.}  In a centralized server scenario, the players must rely on the information coming from the server, since it acts as a mediator, unlike the case of dishonest server, which may be problematic due to erroneous information.
 Furthermore, when money is involved in the games, a server may decide whether or not to transfer money to the player who wins the game.
In addition, a potential hacker may have the opportunity to exploit such loopholes to manipulate the performance of the server, which inherently influences the outcome of the  game.

{\bf Game suspected to a Denial of Service attack (DoS).} At any point in the game, a player who is not satisfied with the score of the game (or for any other reason), has the privilege to launch a denial of service campaign on the server. This is possible, since the server has a single point of failure and there are several low cost service providers for DDoS \cite{DosAS}. An example of DDoS attack that occurred in the wild, can be found in \cite{Dos}.

\subsection{Playing the game over a Blockchain}
A Blockchain architecture that allows arbitrary computation (i.e. {\em Smart contract} \cite{wood2014ethereum}) offers several advantages over a centralized variant, and can mitigate the mentioned flaws from Section \ref{Centrelized}.
\smallskip

\subsubsection{Blockchain benefits}
In this section, we describe the benefits of executing the game over a  Blockchain instead of using centralized server.

{\bf Decentralization.}
A Blockchain that allows Turing-complete computation executes commands across multiple machines, which are called nodes. This architecture enables a trust-less computation and validation over blockchain nodes. This property is in contrast to the case of executing the game on a single server, which makes the Blockchain resilient to denial-of-service attacks. Hence, to "shut down" the computation mechanism, an attacker needs to attack several highly maintained servers across the Internet instead of just a few.

{\bf "The code is the law" paradigm.} Once a smart-contract is uploaded into the blockchain, it cannot be changed\footnote{This claim is not $100\%$ guaranteed, for example in the cases of fork that may arise spontaneously, or with an occurrence of $51\%$ attack, which is rare on a popular Blockchains such as Bitcoin and Ethereum.}. Thus, in a situation where the game is developed with fair rules that can be audited (since the bytecodes are publicly available once uploaded), it becomes infeasible for some entity to interfere and change the rules during an instance of a game.

{\bf Participating in the game cannot be prevented from anyone.}
In addition to the above benefits, playing on  a permissionless Blockchain cannot be censored by a single authority, since every player can create a wallet on their own.

{\bf Instant payment.}
The smart contract code has the ability to transfer money based on certain predetermined programmed rules. Whenever such rules occur, money transfer to a player's wallet cannot be prevented. In our Battleships game, the winner {\em immediately} receives the deposit money of both players (in Ether).

\section{How to enforce fair play?}

\subsection{Survey of possible attacks}\label{attacks}

Building a game over a Blockchain can be mistakenly interpreted to be resilient to cyber attacks. However, such statement tends to be invalid, because a naive implementation would suffer from the following attacks.

{\bf Keeping secrets.} Since the Blockchain maintains a public ledger, putting secret values will expose it to potential cheats, in which an adversary can scan the Blockchain and launch torpedoes on the public locations of his adversary. This is a well known vulnerability that occurs in Blockchain's architecture, more details on this vulnerability can be found in  \cite{atzei2017survey}.

{\bf Dynamically changing battleships' location.} In this attack, the attacker may change the location of the ships dynamically, in his own favor, without updating the other player. Thus, in a condition where there is no enforcement on the location after the first stage of {\em placing the Battleships}, an attacker can attain a major advantage in the game. 

{\bf Inappropriate placement of the battleships.} Using this attack, an attacker can place only a subset of the battleships or the entire battleships such that its parts are neither consecutive or nor forming the correct shape of the ship.

{\bf Implementation's vulnerabilities.} As for any other software, every game could have vulnerabilities and this is specifically more prominent in the logic game executed over blockhcain. A comprehensive survey and taxonomy can be found in \cite{atzei2017survey}.

\smallskip

\subsection{Design concepts}

Herein we describe the design requirements that will mitigate the above attacks. 


{\bf Security requirements.}
\begin{enumerate}

\item  A player that picks a board layout have to commit the board at the beginning of the game, which must not be changed before the game finishes (i.e. a cheater cannot change the location of the battleships without being caught);

\item The above commitment, must not expose any value of the location of the battleships (i.e the locations of the battleships must remain private).

\item The type and size of battleships of the players must obey predefined set of rules (i.e. there must be battleships of sizes $1\times5 , 1 \times 4$, etc.).

\item In each turn, the players need to provide a proof of not cheating about the exact value of the previous torpedo shot toward them, whether it was a {\em Hit} or a {\em Miss}.

\item  Whenever a player makes claim for a victory, he must provide proof that he was not cheated with regard to the location of the battleships.

\item  The game should have a penalty mechanism for a malicious user who is not taking any action at a particular period of time. (i.e the game must prevent the user from freezing the deposit of money in the smart contract due to not continuing the game).


\end{enumerate}

{\bf Architecture requirements.}
\begin{enumerate}

\item The smart contract code of each move should be as light as possible. This requirement is crucial to minimize the finance costs, as well as to provide good user experience.

\item The code should be audited by independent researchers in order to lower the number of implementation's bugs.

\end{enumerate}

\subsection{Game design overview}

This section provides a brief overview of the game design, according to different phases of the game.

\subsubsection{Registration Phase} Two different parties are required to register at the beginning of the game, where both parties make a joint decision on the amount of money to commit to the game. The deposited money is considered a major factor which enforces the players to play by the rules, because any attempt to cheat in the game will result in a punishment of giving the deposited money to the opponent.

\subsubsection{ Placing the battleships}  The players will then choose where to  place their battleships on the board using the game user interface (UI). Afterward, a player who is satisfied with the layout of his own fleet, must upload the computed root of the merkle-tree to the smart contract of the game (in a specified period of time).

{\bf Merkle tree of the board.}
A merkle tree (MT)\cite{merkle1987digital}
 is a cryptographic structure that allows for efficient and secure verification of content. This structure helps to verify the consistency and content of the data. The structure is a binary tree, where every leaf node is labeled with the hash of a data block that it represents and every non-leaf node is labeled with the cryptographic hash of the labels of its child nodes. The topmost node is called the root (similar to a regular binary tree).

\begin{figure}[!htb]

  \begin{minipage}{.35\textwidth}

\begin{tabular}{ |c||c|c| } 
 \hline
 & 1 & 2\\  \hline \hline
 A & X & $\cdot$ \\ \hline
 B& X & $\cdot$ \\  \hline

\end{tabular}

 \end{minipage}%
$\Rightarrow$
\quad \quad
  \begin{minipage}{.5\textwidth}
  
\begin{tikzpicture}[level distance=1.5cm,
  level 1/.style={sibling distance=3cm},
  level 2/.style={sibling distance=1.5cm}]
  \node {Root }
    child {node {A1-A2}
      child {node {A1} child {node {{\bf 2}205}}}
      child {node {A2} child {node {{\bf 0}813}}}
    }
    child {node {B1-B2}
    child {node {B1} child {node {{\bf 2}932}}}
      child {node {B2}  child {node {{\bf 0}417}}}
    };
\end{tikzpicture}

 \end{minipage}%

\caption{Simplified $2 \times 2$ Battleships' board game with only one battleship of size two, and the corresponding merkle-tree structure of that board.}

\label{fig:MT}  
\end{figure}
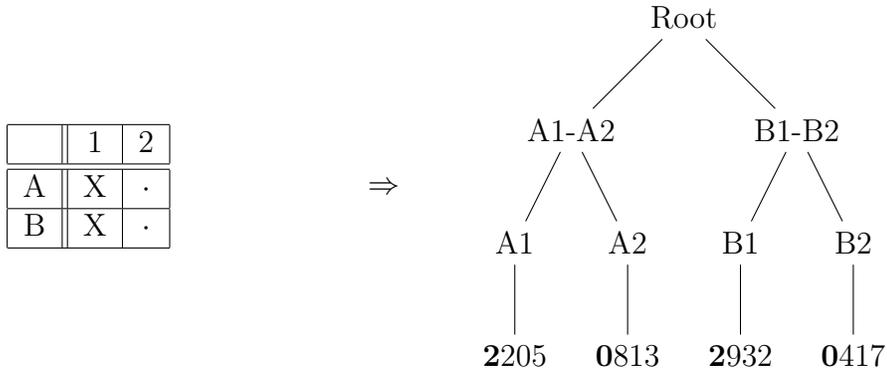

In our design, every leaf node is labeled with a data block in the form of $x||r$, where $x$ denotes whether there exist a ship with size $x$ in the respective cell, or not (i.e $x=0$), and $r$ is a sufficient\footnote{Common length of $r$ can be at least 128 bit, to make it hard to guess the value of the cell by performing Brute-force guessing.} large random value, where the excess number of leafs equals to 0 (those leafs completes the tree to a full binary tree with  $2^{7} =128$ cells).
An illustration of concrete merkle-tree of a simplified board can be seen in Figure \ref{fig:MT}.

 Broadcasting the MT-root will enforce the player to commit the chosen board and force him not to change it later on, since a cryptographic hash function is a one-way function, which is resilient to second-preimage attack\footnote{ The property of second-preimage resistance claims that it is computationally infeasible to find any second input which has the same output as that of a specified input}. Furthermore, when broadcasting this root into the blockhcain, no single value from the underneath values will be revealed, due to the use of random concatenation of each value.

\subsubsection{Launching Torpedoes} In each turn\footnote{Not include the first move.}, a player that wants to launch a torpedo, must broadcast the underneath value of the previous opponent's shot, along with a proof that it is the real value.

The proof is delivered by a MT-path from the MT-root till the targeted cell number of the {\em previous} opponent moves and this path will be verified on the blockchain smart-contract. Only in a situation where the path is valid (i.e. the leaf value fits with the root of the merkle tree uploaded in the first phase), will the player be permitted to perform the next move.
Furthermore, we also enforce time constraint to perform valid moves, in order to avoid a denial of service attack at a particular instance of game.

\subsubsection{Final game verification}

Finally, the smart contract will enforce the candidate winner, which is the player that achieves a correct guess of the entire fleet of the opponent, to reveal his own battleships' locations.
Such rule is necessary to mitigate the {\em Inappropriate placement of battleships attack}.

In case the player refuses to provide a valid MT-paths, he will be tagged as a cheater, and the punishment is that the other player will be announced as the winner, and thus receive the deposited money.


\subsection{Software architecture of the game}
We describe the software architecture of our proposed game design in this section in order to offer a comprehensive overview of the game. The design of the game relies on the 3-tiers architecture \cite{tiers}, which is very similar to a typical decentralized application (dapps).
An illustration of these layers is depicted in Figure \ref{Architecture}.

\begin{enumerate}
\item {\bf Presentation layer} - This layer is responsible for the UI, which includes the following components:

\begin{itemize}
\item  HTML and JScript code that manages the UI of the game. It also includes client side code which ensures game play follow the specified protocol \footnote{This feature is not taken into account in the security analysis, since it is not prevented from malicious attacker who can change the code, and bypass the mechanisms.}

\item Web3.js \cite{Web3} is the layer that connects the HTML client code to interact with the game's smart contract.

\item Metamask wallet\cite{Metamask} enables the users to commit transaction to the blockchain.

\end{itemize}

\item {\bf Logic layer} - This layer is responsible for enforcement of the game rules
and it is placed in the smart contract code. The layer includes the following components:

\begin{itemize}
\item Verification of the boards' MT-path which relies on the solidity library called merkle-tree-solidity \cite{MTP}.

\item Authentication and authorization of the players that participate in the game.

\item Verification of the game rules and validity of transmitted data.

\end{itemize}

\item {\bf Data layer} -This layer is responsible for storing the data that is transferred to the blockchain and includes the following values:

\begin{itemize}
\item The MT-root of the board.
\item The revealed value of cells in the board introduced by previous moves.
\end{itemize}

\end{enumerate}

\begin{figure}[h]
\begin{center}
\label{Architecture}
\begin{tikzpicture}

    \node at (3,3) [yellow] {Presentation layer -\\ {\bf HTML and JScript code}} ;
  
  \node at (3,2) [blue]  {Logic layer-\\  {\bf Battleships smart contract}} ;
  \node at (3,1) [yellow] {Data layer -\\  {\bf  Stored on the blockchain}} ;
    \node at (7,2) [blueM] {MT.sol} ;
    \node at (7,3) [blueM] {Web3.js} ;
    \node at (7,1) [blueM] {Board \\ Committed values} ;

\end{tikzpicture}
\caption{Overview of the architecture's scheme}
\end{center}
\end{figure}

\subsection{Security analysis}
We give a brief security analysis by considering a semi-honest attacker whose computational resources are polynomial bounded. We defer the formal proofs to a full version of this article that will be published in a Journal.

\subsubsection{Security}
This game inherits the basic security mechanism of blockchain which includes:

\begin{itemize}

\item {\em Authentication} of the players during the game will be performed via private key which controls their wallet.

\item {\em Authorization} of performing moves in turns by restricting moves to current player's turn, using smart-contract restrictions.

\end{itemize}

We now proceed to analysis of countermeasures to the types of cheats that were introduced in Section \ref{attacks}.

{\bf Types of cheats.} As discussed previously, any kind of cheat will be punished immediately, by enforcing the rules in the smart contract code. Table \ref{Types} describes cases of potential cheats and how the architecture monitors such cheats in the smart contract.

\begin{center}
\begin{table}[h]

  \begin{tabular}{ | p{50 mm} | p{80 mm} | }
    \hline
    {\bf Type of cheat} & {\bf Countermeasure mechanism }  \\  \hline \hline
     Unresponsive player  & Each turn is time bounded\footnote{A common practice in blockchain is to translate the height of the blockchain to time units, as each block is added in roughly fixed time. Furthermore, to prevent an attacker to perform DDoS of the period of time that the other user makes a move, a full block can be not counted in the number of blocks that needs to be counted.}. 
 \\ \hline
	    Dynamically changing ships & The board is committed via MT-root which stays permanent during a game instance. Any attempt to change the location will produce a fake proof, that the smart contract identifies.\\ \hline
    
Inappropriate placement  & The winner is forced to reveal his fleet before he receives the payment. In case the amount or layout of the battleships does not follow the rules, the player will be punished.\\ \hline

  \end{tabular}
  \caption{Types of cheats and the corresponding countermeasure mechanisms programmed into the smart contract code.}
\label{Types}

\end{table}
\end{center}

\subsubsection{Privacy}
The main privacy issue is how to hide the locations of players' battleships.
Since the entire data in the transactions and smart contracts' fields are public, it must be ensured that they have not exposed parts of  battleships locations which are yet to be made public.
To that end, we examine the messages in each round of the game.

\begin{enumerate}

\item {\bf Commitment phase} - hash function is a one-way function by definition (i.e. given an hash output, it is hard to compute the corresponding input). Thus, an attacker that wants to match boards of size $100$ with the root hash value will have to generate the entire board cells and then compute its MT. 
Since, we concatenate to each block data, a random number with a sufficiently large length, the whole computation complexity is approximately $O({(2^{\lambda}}) ^{100})$, where we denote $\lambda$ as the length of $r$ value in bits and in this experiment we use $\lambda > 128$ bits. Hence, the locations of the boards remain private against a computationally bounded adversary.

\item {\bf Launching torpedoes phase} - in every turn, a player must reveal his targeted board's cell that was threatened by the previous turn. To this end, he broadcast a MT-path from the MT-root till that cell. It is easy to see that the publicly path does not reveal any other intermediate values, which in order to guess them, the attacker needs to generate $2^\lambda$ values, as cryptographic hash function is a one-way function.

\item {\bf Termination phase} - the purpose of this phase is to reveal the candidate's fleet location. Thus, we do not consider any privacy issues in this phase, since the locations are not kept secret at the end of the game.

\end{enumerate}

\subsection{Computational analysis}
This section is concerned with the communication and  computational analyses, which is important to understand the complexity of executing the game, since the computational cost of the game (in Ethereum gas units) is proportional to the number of operations and the data transmitted to the blockchain.
However, we defer the in-depth details of these analyses to the full publication of this article, while we provide here only theoretical analysis.

Let us denote $[H]$, $[B]$, $[BS]$ as the length in bits of the hash function's output, the amount of the cells in the Battleships' board, and the amount of battleships in the game, respectively.

\smallskip

{\bf Communication analysis.}

\begin{itemize}
\item  Commitment phase -both players transmit $[H]$ bytes of the MT-root.

\item Torpedo launching phase - {red}  in each round, the current player transmits MT-path of size $[H] \cdot \log([B])$ and  a cell number of size $\log([B])$ bits.


\end{itemize}

{\bf Computational analysis.}

\smallskip
It is easy to see that the major costly operations derive from the MT proof checking and the termination phase. Thus, the  former computation is bounded by $2[B] \cdot[MTP]$, where $[MTP]$ denotes the cost of executing MT-proof, and the latter computation is bounded by the number of battleships. This is due to the fact that once the valid battleships cells have been received, we only need to check that they are tied to each other.


\section{Future advance mechanisms}
In this section, we present ideas that describe how to extend our proposed game, to include more sophisticated game features and advance game management mechanisms. We also defer the comprehensive description of those features to the full publication version of this article.

\subsubsection{Game variations}

\begin{itemize}

\item {\bf Multi-player case.}
A trivial extension is to simply increase the game to $n$-multi-player game, where in each turn a player will have the privilege to choose the specific board to launch a torpedo toward.

\item {\bf Additional assets.}
In this case, several additional features will be included such as mines, fishes, etc. Players can purchase assets and obtain more rewards for the players. For example, an event of discovering a mine will immediately release a fixed amount of money to the adversary.
These assets can easily be an ERC tokens (both ERC-20 or ERC-721 types).

\end{itemize}

\subsubsection{Game management}

\begin{itemize}

\item {\bf Minimize the cost of moves.} Since each move involves executing transaction on the blockchain, it is desirable to minimize the number of operations to reduce the costs of playing the game. One possible way is to decrease the number of data that is pushed to the main blockchain, by using plasma chains \cite{poon2017plasma}. The latter approach will also increase user usability, since every move will not enforce metamask transaction pop-box.

\item {\bf Enforcing the locations of battleships in advance.}
To enforce that the players' boards is containing exactly (predefined constant) $T$ cells of battleships and that each battleships parts are in a consecutive manner, we can use zero-knowledge proofs. This approach however may increase the communication complexity overhead, which inherently increases the cost of playing the game.

\item {\bf Catching cheaters in advance.}
In our proposed mechanism, despite devising a means for discovering a cheater who tries to change the battleships' locations or sizes and also preventing any form of money theft from the other player at the end of the game, the cheater could still manage to waste the time of the player and postpone the immediate discovery of cheating until the end of the game.

This phenomena occurs since the committed MT-root does not provide a proof of arrangement of the battleships, because the validation of the arrangement only occurs at the end of the game, by the smart contract code. A solution to this problem is to use the zero-knowledge schemes.

In contrast to this, we can transform this "bug" into a feature by  adding a nice "Poker" mechanism feature to the game which allows the players the ability to {\it bluff} each other. As such, a player can choose whether he cheats in advance or not, in the first phase of the game. At any point in time in the game, a player can then guess whether the other player had bluffed him or not.  In case the "cheat" is confirmed, then the player will receive an amount that is inversely proportional to the number of rounds of game already played.

\item {\bf Blacklist of cheaters.}
After detecting cheats in the game, it is possible to take actions against the users that were involved in cheating.
Such actions to those cheaters can be ban them from participating.
However, this feature is in conflict with our requirement to non-censorship game.

\section{Conclusion}
In this article, we proposed the first decentralized Battleships game, which is composed of various cryptographic components to enforce fairness, keep battleships' location secret and protect honest players from malicious cheaters. Furthermore, playing battleships over the blockchain provides major benefits such as making the game DDoS resistant, 
where the money is transferred immediately to the winner, or to the opponent in case cheating is discovered.
The logic of the game is developed using solidity language and deploy over the Ethereum blockchain as a smart contract.

\section{Acknowledgment}
I would like to thank Viki for giving the opportunity to work on this problem. I also want to thank Oded Leiba for the valuable technical discussions, Christiaan Verhoef and Bert Bosman from Amsterdam who showed enthusiasm which encouraged me to refine this game, Polina Zilberman for helping me with proofreading and last but not least, my supervisor Dr. Rami Puzis from the BGU university.

\end{itemize}

\bibliographystyle{plain}
\bibliography{Battleships}
 
 \newpage
 \appendix

\section{Appendix}  

\begin{figure}[h]
\includegraphics[scale=0.25]{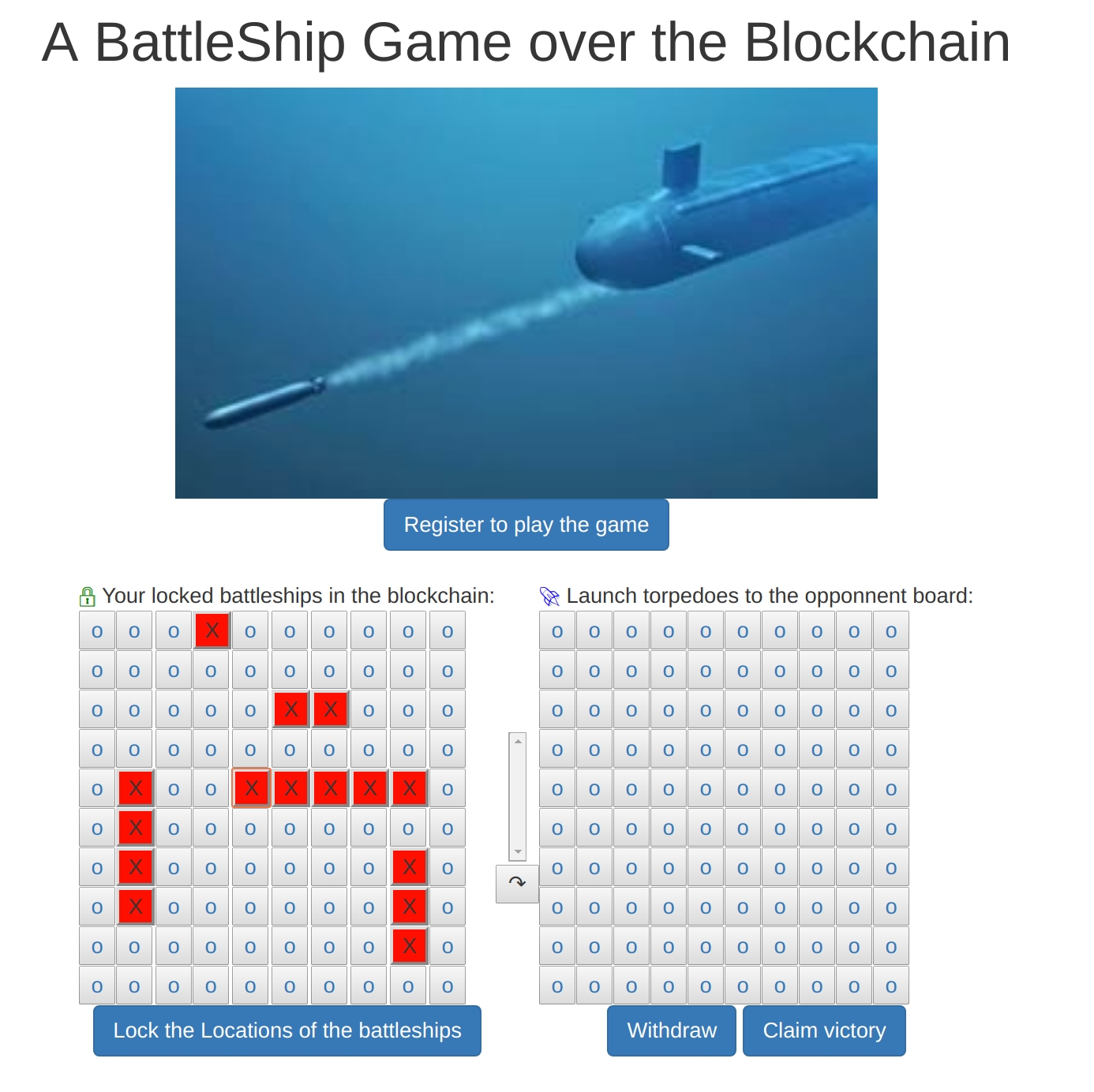}
\caption{A screenshot from the current UI implementation.}

\end{figure}
 
\end{document}